\documentclass[twocolumn,showpacs,preprintnumbers,amsmath,amssymb,10pt]{revtex4}

\usepackage{psfig}
\usepackage{graphicx}
\usepackage{dcolumn}
\usepackage{bm}

\newcommand{\s}{\ensuremath{\psi(t,r)}}

\newcommand{\T}{\ensuremath{\theta}}

\newcommand{\e}{equation} 
\newcommand{\M}{\ensuremath{{\cal M}}}

\newcommand{\X}{\ensuremath{{\cal X}}}

\newcommand{\dw}{\ensuremath{{d\Omega^2_{N-2}}}}

\newcommand{\rdot}{\ensuremath{\dot{R}}}

\newcommand{\F}{\ensuremath{F(r)}}
\newcommand{\f}{\ensuremath{f(r)}}

\begin{document}
\preprint{}
\title{Cosmic Censorship in Higher Dimensions}
\author{Rituparno Goswami}
\email{goswami@tifr.res.in}
\author{Pankaj S Joshi}
\email{psj@tifr.res.in}
\affiliation{ Department of Astronomy and Astrophysics\\ Tata 
Institute of Fundamental Research\\ Homi Bhabha Road,
Mumbai 400 005, India}

\begin{abstract} 
We show that the naked singularities arising in dust collapse
from smooth initial data (which include those discovered by Eardley
and Smarr, Christodoulou, and Newman) are removed when we make 
transition to higher dimensional spacetimes.
The cosmic censorship is then restored 
for dust collapse which will always produce a black hole as the 
collapse end state for dimensions $D\ge6$, under conditions
to be motivated physically such as the smoothness of initial data 
from which the collapse develops.  
\end{abstract}
\maketitle

A considerable debate has continued in recent years
on the validity or otherwise of the cosmic censorship conjecture (CCC)
introduced by Penrose and Hawking, which effectively states that any 
singularities that arise from gravitational collapse from a regular 
initial data must not be visible to far away observers in the spacetime
and are always hidden within black holes.
Such an assumption has been used extensively and is fundamental to the
theory as well as applications of black hole physics.
On the other hand, if a naked singularity results as collapse end 
state, it is no longer necessarily
covered by the event horizon and could communicate, in principle,
with outside observers. Such a scenario is of physical interest because
a naked singularity may have theoretical and observational properties 
quite different from a black hole end state, and communications from 
extreme strong gravity regions dominated by quantum gravity 
may be possible.

Though there is no satisfactory proof or mathematical formulation 
of CCC available despite many efforts, there are many examples of
dynamical collapse models available which lead to a black hole (BH) or
a naked singularity (NS) as the collapse end state, depending on the
nature of the initial data (see e.g.
\cite{rev}
and references therein).
In particular, pioneering analytic models by  Christodoulou,
and Newman, and numerical work by
Eardley and Smarr
\cite{dust}, 
established the existence of {\it shell-focusing} 
naked singularities as the end state of a continual collapse,
where the physical radius of all collapsing shells vanishes. 
In these models the matter form was
taken to be marginally bound dust, assuming that the initial
data functions are smooth and even profiles. 
Newman generalized these models for non-marginally bound class,
thus covering the entire class of dust 
collapse solutions. 
A general treatment for dust collapse with generic initial data 
was developed in
\cite{jd}.

In fact, gravitational collapse of a dust cloud has served, over 
past several decades, as a basic and fundamental paradigm in black
hole physics. This was first discussed by Oppenheimer and 
Snyder 
\cite{osd},
who considered a continual collapse of a homogeneous spherical 
dust cloud in general relativity. The important feature that emerges
is that as the collapse progresses, an event horizon of gravity
forms, which fully covers the spacetime singularity developing
later as the collapse end state. Since the singularity is always covered
by the horizon, we say a black hole has developed as the
collapse final state. This scenario was the basic motivation 
and origin for CCC, stating that any 
physically realistic collapse must
have a similar outcome producing a black hole only.

Our purpose here is to show explicitly, that the above 
mentioned naked singularities of dust collapse
(for both marginally bound and non-marginally bound dust) are removed 
when we go to higher
$(\ge6)$ dimensions, if we allow for only smooth and analytic
initial profiles as assumed by
\cite{dust}. 
In other words, the collapse of a dustlike matter, whose initial 
density and velocity profiles are taken to be smooth
$C^\infty$ functions, which is a condition to be motivated
physically, always goes to a black hole end state for 
dimensions $D\ge6$, and CCC is restored. We point out that the 
increase in dimensions deforms 
the apparent horizon, which is the outer boundary of trapped surfaces, 
in such a manner that the formation of the trapped surfaces 
is advanced. Hence, there exists a critical dimension beyond which 
the neighborhood of 
the center gets trapped before the central singularity and the 
final outcome of collapse is a black hole necessarily.

Such a scenario offers a possibility to recover CCC, because 
it is quite possible that we may be actually living in a universe 
with higher dimensions. The
recent developments in string theory and other field theories 
strongly indicate that gravity is possibly a higher dimensional 
interaction, which reduces to general relativistic description at 
lower energies. Hence, it could be that CCC is restored due 
to extra physical effects arising from our transition itself to 
a higher-dimensional spacetime continuum. 
The recent revival of interest in this problem is
motivated partly by the Randall-Sundrum brane-world
models
\cite{hdim}.

To study the collapse of a spherical dust cloud, we choose 
a general spherically symmetric comoving metric in $N\ge4$ dimensions 
which has the form,
\begin{equation}
ds^2=- e^{\nu(t,r)} dt^2+e^{2\s} dr^2+R^2(t,r) \dw
\label{eq:metric}
\end{equation}
where,
\begin{equation}
\dw=\sum_{i=1}^{N-2}\left[\prod_{j=1}^{i-1}\sin^2(\T^j)\right](d\T^i)^2
\end{equation}
is the metric on $(N-2)$ sphere. The energy-momentum tensor of the dust
has the form,
\begin{equation}
T^t_t=\rho(t,r);\;\;T^r_r=0;\;\;T^{\T^i}_{\T^i}=0
\end{equation}
We take the matter field to satisfy the {\it weak energy condition}, 
that is, the energy density 
measured by any local observer be non-negative, and so for any 
timelike vector $V^i$, we must have,
\begin{equation}
T_{ik}V^iV^k\ge0\;\; \Longrightarrow\;\; \rho\ge0
\end{equation}
In the case of a finite collapsing cloud, there is a finite boundary 
$0<r<r_b$, outside which it is matched to a Schwarzschild exterior. 
The range of the 
coordinates for the metric is then $0<r<\infty$, and $-\infty<t<t_s(r)$
where $t_s(r)$ corresponds to the singular epoch $R=0$.
Solving the $N$-dimensional Einstein equations we get the generalized 
Tolman-Bondi-Lemaitre (TBL) metric 
\cite{TBL}
as,
\begin{equation} 
ds^2= -dt^2 + \frac{R^{'2}}{1+\f}dr^2 + R^2(t,r)\dw
\label{eq:tbl1}
\end{equation}
where $\f$ is an arbitrary function of the comoving radius $r$, 
and $\f>-1$. 
Also we get the required equations of motion as,
\begin{equation}
\frac{(N-2)F'}{2R^{(N-2)}R'}=\rho\;\;\; ; \;\;\; \rdot^2=\frac{\F}
{R^{(N-3)}}+\f
\label{eq:ein5}
\end{equation}
Here $\F$ is another arbitrary function of the comoving
coordinate $r$. 
In spherically symmetric spacetimes 
$\F$ has the interpretation of {\it mass function} which describes 
the mass distribution 
of the dust cloud 
and $\f$ is the {\it energy function}, which specifies the velocity 
profiles for the collapsing shells for any given mass function. The 
energy condition then 
implies $F'\ge0$.
It follows from the above that there is a spacetime singularity
at $R=0$ and at $R'=0$. The latter are called `shell-crossing' 
singularities, which occur when successive shells of matter cross each 
other. These have not been considered generally to be genuine 
spacetime singularities, and possible extensions of spacetime 
have been investigated through the same
\cite{clarke}.
On the other hand, the singularity at $R=0$ is where all matter 
shells collapse to a zero physical radius, and hence this has been 
known as a `shell-focusing' singularity. The nature of this singularity
has been investigated extensively in four-dimensional spacetimes
(see e.g. references in 
\cite{rev}), 
and it is known for
the case of spherical dust collapse that this can be both naked or
covered depending on the nature of the initial data from
which the collapse develops.
The collapse condition is $\dot{R}<0$.

We now use the scaling independence of the comoving 
coordinate $r$ to write
(see e.g. \cite{jd1}),

\begin{equation}
R(t,r)=rv(t,r)
\label{eq:R}
\end{equation}
where,
\begin{equation}
v(t_i,r)=1\;\;\; ;\;\;\; v(t_s(r),r)=0\;\;\; ; \;\;\;\dot{v}<0
\label{eq:v}
\end{equation}
This means we have scaled the coordinate $r$ in such a way that 
at the initial epoch $R=r$, and at the singularity $R=0$. 
It should be noted that we have $R=0$ both at the regular center $r=0$
of the cloud, and at the spacetime singularity, where all matter shells
collapse to a zero physical radius. The regular center is then
distinguished from the singularity by a suitable behaviour of the 
mass function $F(r)$ so that the density remains finite and regular there
at all times till the singular epoch. 
The introduction of the parameter $v$ as above then allows us to 
distinguish 
the spacetime singularity from the regular center, with $v=1$ at 
the initial epoch, including the center $r=0$, which then decreases 
monotonically with time as collapse progresses to value $v=0$ 
at the singularity $R=0$.

In order to ensure the regularity of the initial data, 
it is evident from the equations 
of motion that at 
the initial epoch the two free functions $\F$ and $\f$ must have the 
following forms,
\begin{eqnarray}
\F=r^{(N-1)}\M(r); & \f=r^2b(r)
\label{eq:forms}
\end{eqnarray}
Following
\cite{dust},
we now assume that the initial density and energy functions 
$\rho(r)$ and $f(r)$ are smooth and even, ensuring their analytic 
nature. We note that the Einstein equations as such do not 
impose any such 
restriction, which has to be physically
motivated, and it implies a certain mathematical simplicity
in arguments to deal with a dynamical collapse. 
It follows that both $\M(r)$ and $b(r)$ are now smooth $C^\infty$ 
functions, which means the Taylor expansions of these functions 
around the 
center must be of the following form,
\begin{equation}
\M(r)=\M_0+r^2\M_2+r^4\M_4+\cdots 
\label{eq:forms1}
\end{equation}
and 
\begin{equation}
b(r)=b_0+r^2b_2+r^4b_4+\cdots
\label{eq:forms2}
\end{equation}
that is, all odd terms in $r$ vanish in these expansions,
and the presence of only even terms would ensure smoothness.

To predict the final state of collapse for a given initial
mass and velocity distribution, we shall study below the singularity curve
resulting from the collapse of successive matter shells, and
the apparent horizon developing in the spacetime. A decreasing apparent
horizon in $(t,r)$ plane then is a sufficient condition for a BH as it
shows the entrapment of the neighborhood of the center before the 
central singularity.

Consider now the endless collapse of a dust cloud 
to a final shell-focusing singularity at $R=0$, where
all matter shells collapse to a zero physical radius. 
With the regular initial conditions as above, \e(\ref{eq:ein5}) 
can be written as,
\begin{equation}
v^{\frac{N-3}{2}}\dot{v}=-\sqrt{\M(r)+v^{(N-3)}b(r)}
\label{eq:dotv}
\end{equation}
Here the negative sign implies that $\dot{v}<0$, that is, the matter 
cloud is collapsing. Integrating the above equation with respect to $v$, 
we get,
\begin{equation}
t(v,r)=\int_v^1\frac{v^{\frac{N-3}{2}}dv}{\sqrt{\M(r)+v^{(N-3)}b(r)}}
\label{eq:scurve1}
\end{equation}
We note that the coordinate $r$ is to be treated as a constant in the above 
equation. Expanding $t(v,r)$ around the center, we get,
\begin{equation} 
t(v,r)=t(v,0)+r\X(v)+r^2\frac{\X_2(v)}{2}+r^3\frac{\X_3(v)}{6}+\cdots
\label{eq:scurve2}
\end{equation}
Now from \e(\ref{eq:forms1}) and (\ref{eq:forms2}) we have $\X(v)=0$. 
The function
$\X_2(v)$ is then given as,
\begin{equation}
\X_2(v)=-\int_v^1\frac{v^{\frac{N-3}{2}}(\M_2+v^{(N-3)}b_2)dv}
{(\M_0+v^{(N-3)}b_{0})^{\frac{3}{2}}}
\label{eq:tangent}
\end{equation}
where,
\begin{equation}
b_{0}=b(0) ; \M_0=\M(0); b_2=b''(0); \M_2=\M''(0)
\label{eq:notation}
\end{equation}
We note that the value of $\X_2$ is determined fully
by the initial values of the mass function $\M(r)$ and
the energy function $b(r)$. 
Thus, the time taken for the central shell to reach the singularity is 
given by
\begin{equation}
t_{s_0}=\int_0^1\frac{v^{\frac{N-3}{2}}dv}{\sqrt{\M_0+v^{(N-3)}b_{0}}}
\label{eq:scurve3}
\end{equation}
From the above equation it is clear that for $t_{s_0}$ to be defined,
\begin{equation}
\M_0+v^{(N-3)}b_{0}>0
\label{eq:constraint}
\end{equation}
that is, the continual collapse condition
implies the positivity of the above term.
Hence the time taken for other shells close to the center 
to reach the singularity 
can now be given by,
\begin{equation}
t_s(r)=t_{s_0}+r^2\frac{\X_2(0)}{2}+\cdots
\label{eq:scurve4}
\end{equation}

In order to determine the visibility or otherwise of 
the singularity at $R=0$, we need to analyze the causal 
structure of the trapped surfaces and the nature and behaviour of 
null geodesics in the vicinity of the same. If there exist future
directed null geodesics with past end point at the singularity, which go 
out to faraway observers in spacetime, then the singularity is naked. 
In the case otherwise, we have a black hole resulting as the
end state of a continual collapse.
The boundary of the trapped region of the space-time
is given by the apparent horizon within the collapsing cloud, which
is given by the \e,
\begin{equation}
\frac{F}{R^{N-3}}=1
\label{eq:apphorizon} 
\end{equation}
Broadly, it can be stated that if the neighborhood of the center
gets trapped earlier than the singularity, then it is covered,
otherwise it is naked with
families of non-spacelike future directed trajectories escaping away from 
it. For example, it follows from the above equation that along the
singularity curve $t=t_s(r)$ (which corresponds to $R=0$), for any 
$r>0$ we have $F(r)$ going to a constant positive value, whereas
$R\to 0$. Hence it follows that trapping already occurs before the
singularity develops at any $r>0$ along the singularity 
curve $t_s(r)$.

What we need to determine now is when there will be 
families of non-spacelike paths coming out of the central singularity
at $r=0,t=t_{s}(0)$, reaching 
outside observers, and when there will be none. 
The visibility or otherwise of the singularity is decided
accordingly. By determining the nature of the singularity
curve, and its relation to the initial data, we are able to deduce
whether the trapped surface formation in collapse takes place
before or after the central singularity. It is this causal
structure that determines the possible emergence or otherwise of
non-spacelike paths from the singularity, and settles 
the final outcome in terms of either a BH or NS. 
From \e s(\ref{eq:apphorizon}), we have,
\begin{equation}
v_{ah}(r)=[r^2\M(r)]^{\frac{1}{N-3}}
\label{eq:apphorizon3}
\end{equation}
Using the above equation in (\ref{eq:scurve1}) we get
\begin{equation}
t_{ah}(r)=t_{s}(r)-\int_0^{v_{ah}(r)}\frac{v^{\frac{N-3}{2}}dv}
{\sqrt{\M(r)+v^{(N-3)}b(r)}}
\label{eq:apphorizon4}
\end{equation}
As we are considering the behaviour of the apparent horizon
close to the central singularity at $r=0,R=0$ (all other points $r>0$
on the singularity curve are already covered, as we pointed out 
above), therefore the upper limit of integration 
in the above equation is small, and hence we can expand the integrand
in a power series in $v$, and keep only the leading order term, which 
amounts to,
\begin{equation}
t_{ah}(r)=t_{s_0}+r^2\frac{\X_2(0)}{2}+\cdots - r^{\frac{N-1}{N-3}}
\frac{2}{N-1}\M_0^\frac{1}{N-3}
\label{eq:apphorizon2}
\end{equation}
\begin{figure}[t!!]
\psfig{figure=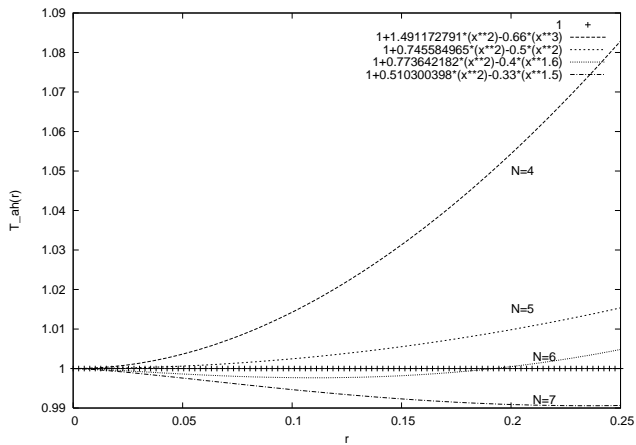,width=8.5cm,angle=-90}
\caption{The apparent horizon in different spacetime dimensions.
Here $\M(r)=1-3.6r^2+\cdots$ and $b(r)=1-3.6 r^2+\cdots$, 
and apparent horizon curves are given for dimensions 4 to 7.}
\end{figure}
Since the apparent horizon is a well-behaved surface for a spherical
dust collapse, hence we can say that the singularity curve
for the collapse and its derivatives around the center are also
well-defined, as the same coefficients are present in both 
(\ref{eq:apphorizon2}) and (\ref{eq:scurve2}).

It is now possible to analyze the effect
of the number of dimensions on the nature and shape of the apparent
horizon. As we increase the number of dimensions and go to dimensions
higher than five,
the negative term in \e (\ref{eq:apphorizon2}) starts dominating, 
thus advancing the trapped surface formation
in time. Fig. 1 shows the behaviour of the apparent horizon curve for
the same initial data in different dimensions.
As seen here, in usual four-dimensional spacetime the above initial
profile ensures an increasing apparent horizon and it can be explicitly
shown that the end state is a NS.
But in six and higher spacetime dimensions we see the negative term  
starts dominating, purely as a result of increase in dimensions.   
This causes the trapped surfaces to form sufficiently early to cover  
the singularity. It follows that the number of spacetime dimensions
causes an intriguing effect on the causal structure and
nature of the apparent horizon, as seen above clearly.

We thus see that for a smooth initial
data and for dimensions higher than five, the apparent horizon becomes
a decreasing function of $r$ near the center. This implies 
that the neighborhood of the center gets trapped before the 
central singularity and the central singularity is then always covered, as 
opposed to the four-dimensional models by
Eardley and Smarr, Christodoulou and Newman, in which case a positive 
$\X_2(0)$ would make the apparent horizon curve increasing and
hence ensures a naked singularity. To be specific, suppose there is
a future directed outgoing null geodesic coming out from the central
singularity at $R=0,r=0$. If $(t_1,r_1)$ is an event along the same, then
$t_1>t_{s_0}$ and $r_1>0$. But for any such $r_1$, the trapped region 
already starts before $t=t_{s_0}$, hence the event $(t_1,r_1)$ is already
in the trapped region and the geodesic cannot be outgoing. Thus,
there are no outgoing paths from the central singularity,
making it covered. It now follows in general (i.e. for both marginally
bound and non-marginal cases) that 
{\it for a dust collapse with smooth initial profiles, the final 
outcome is always a black hole for any spacetime dimensions $D\ge6$, 
and all the naked singularities occurring in  
lower dimensions are removed.}

In five dimensions we have an interesting scenario arising (see also
\cite{deb}). 
As it is clearly seen from the \e (\ref{eq:apphorizon2}), we have a 
critical value of $ \X_2(0)$, below which the apparent horizon is 
decreasing and we will get a BH end state. However, in the case
otherwise, a naked singularity can result. 
It is interesting to note also that the above results holds
if the initial profiles, instead of being absolutely smooth
$C^\infty$ functions, are taken to be sufficiently smooth
({\it i.e.} at least a $C^2$ function).

We note that it may still be possible to have both BH/NS
outcomes as collapse end states when initial data and metric are
not assumed necessarily to be $C^\infty$ or at least $C^2$ analytic functions
\cite{jd}.
Hence such smoothness conditions need to be investigated and
probed further carefully to see if they could be strongly motivated 
from a physical perspective. 
It will also be interesting to see the effects of introducing 
pressures within the collapsing cloud, and to examine if we could
still avoid naked singularities and preserve cosmic censorship
in higher dimensions by introducing physically motivated conditions
such as above. This will be discussed elsewhere.

\end{document}